\documentclass[10pt]{article}
\usepackage{amssymb}
\usepackage{verbatim}
\renewcommand{\theequation}{\thesection-\arabic{equation}}
\makeatletter
\@addtoreset{equation}{section}

\catcode`\@=11

\newcounter{app}

\def\app{\setcounter{equation}{0}
\def\theequation{A\arabic{app}.\arabic{equation}}\par
   \addvspace{4ex}
   \@afterindentfalse
  \secdef\@app\@dapp}
\newcommand\@app{\@startsection {app}{1}{0ex}%
                                   {-3.5ex \@plus -1ex \@minus -.2ex}%
                                   {2.3ex \@plus.2ex}%
                                   {\normalfont\Large\bf}}

\def\@dapp#1{%
{\parindent \z@ \raggedright  \bf #1}\par\nobreak}
\def\l@app#1#2{\ifnum \c@tocdepth >\z@
    \addpenalty\@secpenalty
    \addvspace{1.0em \@plus\p@}%
    \setlength\@tempdima{8.5em}%
    \begingroup
      \parindent \z@ \rightskip \@pnumwidth
      \parfillskip -\@pnumwidth
      \leavevmode \bfseries
      \advance\leftskip\@tempdima
      \hskip -\leftskip
      #1\nobreak\hfil \nobreak\hb@xt@\@pnumwidth{\hss #2}\par
    \endgroup\fi}
\catcode`\@=12

\def\be{\begin{equation}}
\def\ee{\end{equation}}
\def\bea{\begin{eqnarray}}
\def\eea{\end{eqnarray}}

\renewcommand{\l}{\langle}
\renewcommand{\r}{\rangle}

\newtheorem{lemma}{Lemma}[section]
\newtheorem{remark}{Remark}[section]

\newtheorem{proposition}{Proposition}[section]

\def\bprop{\begin{proposition}\rm}
\def\eprop{\end{proposition}}
\def\blemm{\begin{lemma}\small}
\def\elemm{\end{lemma}}
\def\bremark{\begin{remark}}
\def\eremark{\end{remark}}
\def \tr{{\rm tr}}
\begin{document}
\begin{flushright}
CRM-2907 (2002)\\
nlin.SI/0211051
\end{flushright}
\vspace{0.2cm}
\begin{center}
\begin{Large}
\textbf{Scalar products of symmetric functions
and matrix integrals}\footnote{
Work supported in part by the Natural Sciences and Engineering Research
Council of Canada (NSERC) and the Fonds FCAR du Qu\'ebec, the
LDRD project 20020006ER ``Unstable fluid/fluid interfaces'' at Los Alamos
National Laboratory and grant RFBR02-02-17382a.}
\end{Large}\\
\vspace{1.0cm}
\begin{large} {J. Harnad }$^{\dagger
\ddagger}$\footnote{harnad@crm.umontreal.ca} and
 { A. Yu. Orlov}$^{\star}$\footnote{orlovs@wave.sio.rssi.ru}
\end{large}
\\
\bigskip
\begin{small}
$^{\dagger}$ {\em Centre de recherches math\'ematiques,
Universit\'e de Montr\'eal\\ C.~P.~6128, succ. centre ville, Montr\'eal,
Qu\'ebec, Canada H3C 3J7} \\
\smallskip
$^{\ddagger}$ {\em Department of Mathematics and
Statistics, Concordia University\\ 7141 Sherbrooke W., Montr\'eal, Qu\'ebec,
Canada H4B 1R6} \\
\smallskip
$^{\star}$ {\em Nonlinear Wave Processes Laboratory, \\
Oceanology Institute, Nakhimovskii Prospect 36 \\
Moscow 117851, Russia} \\
\end{small}
\bigskip
\bigskip
{\bf Abstract}
\end{center}
\begin{small}

We present relations between Hirota-type bilinear operators, scalar
products on spaces of symmetric functions and integrals defining matrix model
partition functions. Using the fermionic Fock space representation, a proof of
the expansion of an associated class of KP and 2-Toda tau functions $\tau_{r,n}$
in a series of Schur functions generalizing the hypergeometric series is
given and related to the scalar product formulae. It is shown how special cases
of such $\tau$-functions may be identified as formal series expansions of
partition functions. A closed form exapnsion of $\log \tau_{r,n}$ in terms of
Schur functions is derived.

\end{small}


\section{Partitions, Schur functions and scalar products }

\subsection{Partitions}

We recall here some standard definitions and notations concerning
symmetric functions and partitions (see \cite{Mac} for further
details). Symmetric polynomials of many variables are
parameterized by partitions. A {\em partition} is any (finite or
infinite) sequence of non-negative integers in decreasing order:
\be \label{partition} \lambda = (\lambda_1, \lambda_2,
\dots,\lambda_r,\dots ),\quad \lambda_1 \ge \lambda_2 \ge \dots
\ge\lambda_r \ge \dots \ee The nonzero $\lambda_i$'s in
(\ref{partition}) are called the {\em parts} of $\lambda$; the
number of parts is the {\em length} of $\lambda$, denoted
$l(\lambda)$, and the sum of the parts is the {\em weight} of
$\lambda$, denoted by $|\lambda|$. If $n=|\lambda|$ we say that
$\lambda$ is a {\em partition of} $n$. The partition of zero is
denoted $(0)$.

The {\em diagram} of a partition (or {\em Young diagram}) may be
defined as the set of points (or nodes) \hbox{$\{(i,j) \in {\bf Z}^2\}$} such
that
$1\le j \le \lambda_i$. Thus a Young diagram may be viewed as a subset
of entries in a matrix with $l(\lambda)$ rows and $\lambda_1$
columns. We shall denote the diagram of $\lambda$ by the same symbol.

For example

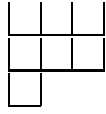
\vbox{
\qquad\qquad\qquad\qquad\qquad\qquad\qquad\begin{tabular}{|c|}
   \\ \hline
\end{tabular}\begin{tabular}{|c|}
   \\ \hline
\end{tabular}\begin{tabular}{|c|}
   \\ \hline
\end{tabular}

\qquad\qquad\qquad\qquad\qquad\qquad\qquad\begin{tabular}{|c|}
   \\ \hline
\end{tabular}\begin{tabular}{|c|}
   \\ \hline
\end{tabular}\begin{tabular}{|c|}
   \\ \hline
\end{tabular}

\qquad\qquad\qquad\qquad\qquad\qquad\qquad\begin{tabular}{|c|}
   \\ \hline
\end{tabular}\\}
\noindent
is the diagram of $(3,3,1)$. The weight of this partition is $7$
and the length is $3$.

Another notation for partitions is due to Frobenius. Suppose that
the main diagonal of the diagram of $\lambda$  consists of $r$
nodes $(i,i)\quad (1\le i\le r)$. Let $\alpha_i=\lambda_i-i$ be
the number of nodes in the $i$th row of $\lambda$ to the right of
$(i,i)$, for $1\le i\le r$, and let $\beta_i=\lambda_i'-i$ be the
number of nodes in the $i$th column of $\lambda$ below $(i,i)$,
for $1\le i\le r$. We have $\alpha_1>\alpha_2>\cdots >\alpha_r\ge
0$ and $\beta_1>\beta_2>\cdots >\beta_r\ge 0$. In Frobenius'
notation the partition $\lambda$ is denoted \be \label{Frob}
\lambda = \left( \alpha_1,\dots ,\alpha_r|\beta_1,\dots
,\beta_r\right)=(\alpha |\beta ) \ . \ee This corresponds to a
hook decomposition of the diagram of $\lambda$, where the biggest
hook is $\left( \alpha_1|\beta_1\right)$, the next is
$\left(\alpha_2|\beta_2\right)$, and so on, down to the smallest
one, which is $\left( \alpha_r|\beta_r\right)$. The corners of the
hooks are situated on the main diagonal of the diagram. For
instance the partition $(3,3,1)$ consists of two hooks $(2,2)$ and
$(1,0)$, \break

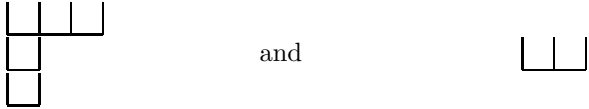
\vbox{
\qquad\qquad\begin{tabular}{|c|}
   \\ \hline
\end{tabular}\begin{tabular}{|c|}
   \\ \hline
\end{tabular}\begin{tabular}{|c|}
   \\ \hline
\end{tabular}

\qquad\qquad\begin{tabular}{|c|}
   \\ \hline
\end{tabular}
\qquad\qquad\qquad\qquad and
\qquad\qquad\qquad\qquad\begin{tabular}{|c|}
   \\ \hline
\end{tabular}\begin{tabular}{|c|}
   \\ \hline
\end{tabular}

\qquad\qquad\begin{tabular}{|c|}
   \\ \hline
\end{tabular}
\\}
\noindent
so, in Frobenius notation this is $(2,1|2,0)$.

\subsection {Scalar products of Schur functions}

 In applications to KP theory it is useful to view the Schur function
$s_\lambda$ as a basis in the space of weighted homogeneous polynomials in an
infinite sequence of parameters ${\bf \gamma}=(\gamma_1,\gamma_2,\dots )$ (with
weight$(\gamma_j)=j$) defined by
\be
\label{Schurht}
s_{{\lambda}}({\bf \gamma})=\det(h_{\lambda_i-i+j}({\bf
\gamma}))_{1\le i,j\le r} ,
\ee
where $\{h_n({\bf \gamma})\}_{n\in {\bf Z}}$ are the elementary Schur
functions (or complete symmetric functions) defined by the Taylor expansion
\be
\label{elSchur}
e^{\xi({\bf \gamma},z)}: =\exp(\sum_{k=1}^{\infty}\gamma_kz^k)=
\sum_{n=0}^{\infty}z^n h_n({\bf \gamma})  \ ,
\ee
and $h_n({\bf \gamma}):=0$ if $n<0$.
If the parameters ${\bf \gamma}=(\gamma_1, \gamma_2, \dots )$ are
determined in terms
 of a finite number of variables $(x_1, \dots x_N)$ by
 \be \gamma_j \label{Miwa}= {1\over
j}\sum_{a=1}^N x_a^j   \ ,
\ee
i.e., if the Schur functions are interpreted as irreducible characters for
$GL(N)$), we use the notation $s_\lambda([{\bf x}])$, where \be [{\bf x} ]:=
(\sum_{a=1}^N x_a, \sum_{a=1}^N x^2_a, \dots )  \ .
\ee

The Cauchy-Littlewood identity \cite{Mac} provides a generating function 
formula for the full set of Schur functions, viewed as functions of two finite
sets of variables $(x_1, \dots x_N)$, $(y_1, \dots y_M)$:
\be
\prod_{a=1}^N \prod_{b=1}^M (1-x_a
y_b)^{-1} = \sum_\lambda s_\lambda({\bf t})s_\lambda({\bf t}^*) \ ,
\label{Cauchy}
\ee
where
\be
t_j = {1\over j}\sum_{a=1}^N  x_a^j, \qquad t^*_j =
 {1\over j}\sum_{a=1}^M
y_a^j  \ ,
\label{finiteparam}
\ee
and ${\bf t}:=(t_1,t_2,\dots)$ and  ${\bf t}^*:=(t_1^*,t_2^*,\dots)$. 

A scalar product \cite{Mac} is defined on the space of weighted homogeneous
polynomials in an infinite sequence of variables, such that the Schur functions
are orthonormal
\be
<s_\lambda ,s_\mu >=\delta_{\mu,\lambda} \ .
\label{stand}
\ee

In the context of the KP and Toda lattice hierarchies the variables $\{x_i\}$
in (\ref{Miwa})  (possibly with $N=\infty$) are called the Hirota-Miwa variables
while $\gamma=(\gamma_1,\gamma_2,\dots)$ play the role of KP flow parameters.

\section{A deformation of the standard scalar product and series
of hypergeometric type}

\subsection{Scalar product $<,>_{r,n}$}

Given a function $r(n)$ defined on the integers $n\in {\bf Z}$ and  a partition
$\lambda$ of weight $|\lambda|$ and length $l(\lambda)$, let us define
\be
\label{rlambda} r_\lambda(n):=\prod_{i,j\in \lambda}r(n+j-i)
=\prod _{i=1}^{l(\lambda)} \prod_{j=1}^{\lambda_i}r(n+j-i) .
\ee
Thus  $r_\lambda(n)$ is a product of the values of the function $r$
translated by the {\it content}  ($j-i$) of the node, taken over all nodes of
the Young diagram of the partition $\lambda$. For instance, for the partition
$(3,3,1)$,
\be
r_{(3,3,1)}(n)=r(n+2)(r(n+1))^2(r(n))^2r(n-1)r(n-2) \ .
\ee
For the zero partition one puts $r_0 := 1$.

Given a function $r$ and an integer $n$ which is greater than or equal to the
largest zero of $r$,  we may define an associated bilinear form
$<\  ,\  >_{r,n}$ as follows:
\be
\label{scalrn}
<s_\lambda,s_\mu>_{r,n}=r_\lambda(n)\delta_{\lambda\mu}  \ .
\ee
Let $\{n_i \in Z\}$ be the zeros of $r$ and
\be
\label{k}
k:=\min |(n-n_i)|.
\ee
 The product (\ref{scalrn}) is nondegenerate on the space $\Lambda_k$ of
symmetric polynomials in $k$ variables ${\bf x}^k:=(x_1, \dots x_k)$. This
follows from the fact that, from the definition (\ref{rlambda}), the quantity
$r_\lambda (n)$ never vanishes for  partitions $\lambda$ with length $l(\lambda)
\le k$. In this case the Schur functions of $k$ variables $\{s_\lambda([{\bf
x}^k]), l(\lambda)\le k\}$ form a basis for $\Lambda_k$. If, on the contrary,
$n-n_i<0$ for all zeros of $r$, then the factor $r_\lambda(n) $ never
vanishes for the partitions $\{\lambda :l(\lambda')\le k\}$ , where $\lambda'$
is the conjugate partition, and
$\{s_\lambda({-[\bf x}^k]), l(\lambda')\le k\}$ form a basis on $\Lambda_k$.
 If $r$ is a nonvanishing function then the scalar product is non-degenerate
on $\Lambda_\infty$.

\subsection{The main example.}

The case $r(n)=n$ plays a special role in applications to
two-matrix models. We denote by $<,>_n'$ the scalar product for
this case 
\be 
\label{<>'} <s_\lambda
,s_\mu>_n'=(n)_\lambda \delta_{\lambda \mu}, \quad
(n)_\lambda=\prod_{i=1}^{l(\lambda)}\prod_{j=1}^{\lambda_i}
(n-i+j) \ . 
\ee 
(Recall that the restriction of this to any space of symmetric polynomials
in $\le n$ variables is non-degenerate.)
 For a pair $(f,g)$ of symmetric functions of $n$ variables ${\bf
x}=(x_1,\dots,x_n)$, we have the following simple realization of
this scalar product.
\be
\label{difopr=n}
<f\, ,g\,>'_n=
\frac{1}{c_n}\Delta(\partial)f(\partial)\cdot \Delta({\bf
x})g({\bf x})|_{{\bf x}=0} \ ,\quad c_n=\prod_{k=1}^n k!
\ee
where $\Delta({\bf x}):=\prod_{1\le i<j\le n}(x_i-x_j)$ is the Vandermonde
determinant,
\be
\Delta(\partial):=\prod_{1\le i<j\le
n}\left(\frac{\partial}{\partial x_i}-\frac{\partial}{\partial
x_j}\right)  \ ,
\ee
and $f(\partial)$ is the operator obtained by replacing   $\{x_i\}$
 by $\{\frac{\partial}{\partial x_i}\}$. The proof follows from the
Jacobi-Trudi formula for the Schur function \cite{Mac}
\be
s_\lambda({\bf x})= \det
\left({x_i}^{\lambda_j+n-j}\right)_{i,j=1,\dots,n}/(\Delta({\bf x}) \ ,
\ee
and the formula
\be
\label{r=kschur} \frac{1}{c_n}\det
\left(\partial_{x_i}^{\lambda_j+n-j}\right)_{i,j=1,\dots,n}\cdot
\det \left({x_i}^{\mu_j+n-j}\right)_{i,j=1,\dots,n}|_{{\bf x}=0}=
(n)_{\lambda} \delta_{\lambda\mu}=<s_\lambda,s_\mu>_{r,n}
\ee
(It is interesting to compare the scalar product (\ref{difopr=n})
with the one considered in \cite{VF}.)

We present two different realizations of the formula
(\ref{difopr=n}) via multiple integrals.

(A)
\be
\label{realizA}
<f,g>_{n}'=\frac{1}{\pi^n c_n}\int_C \cdots \int_C f({\bf
z})g({\bar {\bf z}})e^{-|z_1|^2-\cdots -|z_n|^2}|\Delta({\bf
z})|^2 d^2 z_1 \cdots d^2 z_n  \ ,
\ee
where the integration is over $n$ copies  $\{z_i\}_{i=1\dots n}$ of the complex
plane.

The proof of (\ref{realizA}) follows from the following relation, valid for any
pair of polynomial functions $(a,b)$ of one variable
\be
\label{fgintA}
a\left(\frac{\partial}{\partial z}\right)\cdot b(z)|_{z=0}=
\frac{1}{\pi}\int_C a(z)b({\bar {z}})e^{-|z|^2}dz d{\bar z}  \ .
\ee

(B) \be \label{realizB} <f,g>_{n}'={1 \over {(2\pi i)^n c_n}}\int_\Re\int_\Im
\cdots
\int_\Re\int_\Im f({\bf x})g( {\bf y})e^{-(x_1y_1+ \cdots +x_ny_n})\Delta({\bf
x})\Delta({\bf y}) d x_1dy_1 \cdots d x_n dy_n \ , \ee where the
integration is over $n$ copies $\{x_i\}_{i=1\dots n}$ of the real
line and $n$ copies $\{y_i\}_{i=1\dots n}$ of the imaginary line.
 If $f$ and $g$ are polynomials, these integrals must be evaluated in the sense
of distributions, with a suitable regularization procedure. Alternatively, we
may interpret (\ref{realizB}) as applied to a pair of functions $f, g$ that
decrease rapidly enough at $\infty$ to make the integral converge.

The proof of (\ref{realizB}) follows from the formula 
\be
\label{fgintB} a\left(\frac{\partial}{\partial x}\right)\cdot
b(x)|_{x=0}= {1\over 2\pi i}\int_\Re\int_\Im a(
x)b({y})e^{- xy}dxdy \ . 
\ee

Below we shall mainly be concerned with the following series \be
\label{taurhatxx} \frac{1}{c_n}e^{\sum_{m=0}^\infty\sum_{i=1}^n
t_m\partial_{x_i}^m}\Delta(\partial)\cdot
e^{\sum_{m=0}^\infty\sum_{i=1}^n t_m^*{x}_i^m}\Delta({x})|_{{\bf
x}=0}=\sum_{{\lambda}} (n)_\lambda s_{\lambda }({\bf t})s_{\lambda
}({\bf t^* }) \ . \ee This equality follows from (\ref{<>'}),
(\ref{difopr=n}) and the generalized Cauchy-Littlewood identity
(see (\ref{Schurdoublegener}) below). The r.h.s. is generally
divergent, and must be interpreted  as a formal graded polynomial
series. It may be identified with the hypergeometric series
${_2{\mathcal{F}}}_0\left(n,n ;{\bf t}, {\bf t}^* \right)$, see
\cite{OS}.

\section{Fermionic operators, vacuum expectations,
tau functions}

\subsection{ Free fermions}
In this section we use the formalism of fermionic Fock space as in
\cite{DJKM}. The algebra of {\em free fermions} is the infinite
dimensional Clifford algebra ${\bf A}$ over ${\bf C}$ with
generators $\psi_n,\psi_n^* (n \in Z)$ satisfying the
anti-commutation relations: \be
 \label{antikom}
[\psi_m,\psi_n]_+=[\psi^*_m,\psi^*_n]_+=0;\qquad [\psi_m,\psi^*_n]_+=
\delta_{mn} \ , \quad m,n \in {\bf Z} \ .
\ee
Any element of $W=\left(\oplus_{m \in Z}{\bf C}\psi_m\right)\oplus
\left(\oplus_{m\in Z}{\bf C}\psi_m^*\right)$
 will be referred to  as a {\em free fermion}. The Clifford algebra has a
standard representation ({\em Fock representation}) as follows. Put
$W_{an}=\left(\oplus_{m<0}{\bf C}\psi_m\right)\oplus
\left(\oplus_{m\ge 0}{\bf C}\psi_m^*\right)$, and
$W_{cr}=\left(\oplus_{m\ge 0}{\bf C}\psi_m\right)\oplus
\left(\oplus_{m< 0}{\bf C}\psi_m^*\right)$, and consider the left
(resp. right) ${\bf A}$-module $F={\bf A}/{\bf A}W_{an}$ (resp
$F^*=W_{cr}{\bf A}{\backslash}{\bf A}$). These are cyclic ${\bf
A}$-modules generated by the vectors $|0\r= 1$ mod ${\bf A}W_{an}$
(resp. by $\l 0|= 1$  mod $W_{cr}{\bf A}$), with the properties
\begin{eqnarray}\label{vak}
\psi_m |0\r=0 \qquad (m<0),\qquad \psi_m^*|0\r =0 \qquad (m \ge 0) , \\
\l 0|\psi_m=0 \qquad (m\ge 0),\qquad \l 0|\psi_m^*=0 \qquad (m<0) .
\end{eqnarray}
Vectors $\l 0|$ and $|0\r$ are referred to as left and right
vacuum vectors. Fermions $w\in W_{an}$ annihilate the left vacuum
vector, while fermions $w\in W_{cr}$ annihilate the right vacuum
vector.

The Fock spaces $F$ and $F^*$ are mutually dual, with the pairing
defined through the linear form $\l 0| |0 \r$ on ${\bf A}$
called the {\em vacuum expectation value}. This is given by
\begin{eqnarray}\label{psipsi*vac}
\l 0|1|0 \r=1,\quad \l 0|\psi_m\psi_m^* |0\r=1\quad m<0,
\quad  \l 0|\psi_m^*\psi_m |0\r=1\quad m\ge 0 ,
\end{eqnarray}
\be
\label{end}
 \l 0|\psi_m\psi_n |0\r=\l 0|\psi^*_m\psi^*_n |0\r=0,\quad \l
0|\psi_m\psi_n^*|0\r=0 \quad m\ne n ,
\ee
and by {\bf the Wick rule}, which is
\be
\label{Wick}
\l 0|w_1 \cdots w_{2n+1}|0 \r =0,\quad \l 0|w_1 \cdots w_{2n} |0\r
=\sum_\sigma sgn\sigma \l 0|w_{\sigma(1)}w_{\sigma(2)}|0\r \cdots
\l 0| w_{\sigma(2n-1)}w_{\sigma(2n)} |0\r ,
\ee
where $w_k \in W$, and $\sigma$ runs over permutations such that
$\sigma(1)<\sigma(2)\dots  \sigma(2n-1)<\sigma(2n)$ and
$\sigma(1)<\sigma(3)<\cdots <\sigma(2n-1)$.

\subsection{The Lie algebra $\widehat {gl}(\infty)$}
Consider infinite matrices $(a_{ij})_{i,j\in Z}$ satisfying the condition that
there exists an $N$ such that $a_{ij}=0$ for $|i-j|>N$. Such matrices
are called generalized Jacobi (or ``finite band'') matrices, and  form a Lie
algebra under the usual matrix commutator bracket.

Let $:\dots :$ denote the normal ordering operator, defined such that 
\be
:\psi_i\psi_j^*:=\psi_i\psi_j^*-\l 0|\psi_i\psi_j^*|0\r . \ee
The space of linear combinations of quadratic elements of the form
\be
\sum_{i, j \in {\bf Z}} a_{ij}:\psi_i\psi_j^*:  \ ,
\ee
together with the identity element $1$,
span the infinite dimensional Lie algebra $\widehat{gl}(\infty)$:
\be
\label{commutator}
[\sum a_{ij}:\psi_i \psi_j^*:,\sum b_{ij} :\psi_i\psi_j^*:]=
\sum c_{ij}:\psi_i \psi_j^*:+c_0  \ .
\ee
where
\be
 c_{ij}=\sum_k a_{ik}b_{kj}-\sum_k b_{ik}a_{kj} ,
\ee
The last term
\be
\label{cocycle}
c_0=\sum_{i<0,j\ge 0} a_{ij}b_{ji}-\sum_{i\ge 0,j<0} a_{ij}b_{ji} .
\ee
is central, so the  Lie algebra $\widehat {gl}(\infty)$ is a central
extension of the algebra of generalized Jacobian matrices.

\subsection{ Bilinear identity}
Let  $g$ be the exponential of an operator in
$\widehat{gl}(\infty)$:
\be
\label{expLie}
g=\exp \sum_{i,j \in {\bf Z}} a_{ij}:\psi_i\psi_j:  \ .
\ee
i.e., an element of the corresponding group.
Using (\ref{commutator}) it is possible to derive the following
relation
\be
 g\psi_n=\sum_m \psi_m A_{mn}g \ , \qquad \psi^*_n g=
g\sum_m A_{nm}\psi^*_m  \ , \label{rot} \ee where the coefficients
$A_{nm}$ are determined by $a_{nm}$.  In turn (\ref{rot}) implies
\cite{DJKM} \be
 \label{bilinear}
[\sum_{n \in Z}\psi_n \otimes \psi_n^*, g \otimes g ]=0  \ .
\ee
This last relation is very important for applications to integrable systems,
and is equivalent to the  Hirota bilinear equations.

\subsection{Double KP flow parameters. The KP and TL tau functions.}

We introduce the following operators.
\be
\label{hamiltonians}
H_n=\sum_{k=-\infty}^\infty \psi_k\psi^*_{k+n},\quad n\neq 0,
\quad
 H({\bf
t})=\sum_{n=1}^{+\infty} t_n H_n ,\quad H^*({\bf
t}^*)=\sum_{n=1}^{+\infty} t_n^* H_{-n}.
\ee
Here,  $H_n \in \widehat{gl}(\infty)$, and $H({\bf t}),H^*({\bf t}^*)$
also belong to $\widehat{gl}(\infty)$ if we restrict the number of
non-vanishing parameters $\{t_m,t_m^*\}$ to be finite. For $H_n$ we have the
Heisenberg algebra commutation relations:
\be
\label{Heisenberg}
[H_n,H_m]=n\delta_{m+n,0} .
\ee
Note also that
\be
\label{Hvac}
H_n|0\r=0=\l 0|H_{-n},\quad n>0 \ .
\ee

For what follows, we also need to introduce the free fermion field operators
\be
 \label{fermions}
\psi(z):=\sum_k \psi_k z^k ,\qquad \psi^*(z):=\sum_k \psi^*_k
z^{-k-1}dz \ .
\ee

 For each $n\in {\bf N}^+$, define the ``charge $n$'' vacuum vector
  labelled by the integer $n$:
\begin{eqnarray}
\l n|:=\l 0|\Psi^{*}_{n},\qquad |n\r :=\Psi_{n}|0\r ,
\end{eqnarray}
\begin{eqnarray}
\Psi_{n}=\psi_{n-1}\cdots\psi_1\psi_0 \quad n>0,
\qquad \Psi_{n}=\psi^{*}_{n}\cdots\psi^{*}_{-2}\psi^{*}_{-1}\quad n<0 ,
 \nonumber\\
\Psi^{*}_{n}=\psi^{*}_{0}\psi^{*}_1\cdots\psi^{*}_{n-1} \quad n>0,
\qquad \Psi^{*}_{n}=\psi_{-1}\psi_{-2}\cdots\psi_{n}\quad n<0 .
\end{eqnarray}

Given any $g$ satisfying the bilinear identity (\ref{bilinear}),
 the corresponding KP and two dimensional TL tau-functions
 \cite{DJKM} are
defined as follows:
\be
\label{taucorKP}
\tau_{KP}(n,{\bf t}):= \langle n|e^{H({\bf t})}g|n \rangle ,
\ee
\be
\label{taucor}
\tau_{TL}(n,{\bf t},{\bf t}^*):= \langle n|e^{H({\bf
t})}ge^{H^*({\bf t}^*)}|n \rangle .
\ee

The first set of parameters appearing in $\tau_{KP}(n, {\bf t})$ are the usual
KP flow parameters, while the parameters ${\bf t}=(t_1,t_2,\dots)$ and 
${\bf t}^*=(t_1^*,t_2^*,\dots)$ in $\tau_{TL}(n, {\bf t}, {\bf t}^*)$
are sometimes called  the higher (two dimensional) Toda lattice
times \cite{DJKM, UT}. (If we fix $n$ and the parameters ${\bf t^*}$, then $\{\tau_{TL}(n, {\bf
t} , {\bf t}^*)\}_{n\in {\bf N}}$ may also be viewed as KP tau-functions
in the ${\bf t}$ variables.)
The first  three parameters $(t_1,t_2,t_3)$ are just
the independent variables  appearing in the KP equation \cite{ZM}
\be
 4\partial_{t_1}\partial_{t_3}u=\partial_{t_1}^4 u+
3\partial_{t_2}^2u+3\partial_{t_1}^2 u^2 \quad
(u=2\partial_{t_1}^2\log \tau) \ .
\ee

\section{KP tau-function $\tau_r(n,{\bf t},{\bf t^*} )$}

\subsection{Schur function expansions.}
For each choice of the parameters ${\bf t^*} =(t_1^*,t_2^*,\dots)$ we
define the operator
\be
\label{Abeta}
A({\bf t^*}) :=\sum_{m=1}^\infty t_k^*A_k  \ ,
\ee
where the $A_k$'s, which are defined as
\be
\label{A_k} A_k=
\sum_{n=-\infty}^\infty \psi^*_{n-k} \psi_n r(n)r(n-1)\cdots
r(n-k+1), \quad k=1,2,\dots \ ,
\ee
commute amongst themselves
\be
[A_k, \, A_l] = 0,  \qquad \forall j, k \ .
\ee
Using the explicit form of $A_k$ and the anti-commutation relations
(\ref{antikom}) we obtain
\be
\label{psixir}
e^{A({\bf t^*})}\psi(z)e^{-A({\bf t}^*)}= e^{-\xi_r({\bf
t}^*,z^{-1})}\cdot \psi(z)
\ee
\be
\label{psi*xir}
e^{A({\bf t}^*)}\psi^*(z)e^{-A({\bf t}^*)}= e^{\xi_{r'}({\bf
t}^*,z^{-1})}\cdot \psi^*(z)  \ ,
\ee
where $\xi_r({\bf t}^*,z^{-1})$ and $\xi_{r'}({\bf t}^*,z^{-1})$ are operators
on functions of the auxiliary parameter $z$, defined by
\be
\label{xir}
\xi_r({\bf t}^*,z^{-1}) :=\sum_{m=1}^{+\infty}t_m\left(\frac 1z
r(D)\right)^m, \qquad
\xi_{r'}({\bf t}^*,z^{-1}) :=\sum_{m=1}^{+\infty}t_m\left(\frac 1z
r'(D)\right)^m \ ,
\ee
with
\be
  D:=z\frac{d}{dz},\qquad r'(D):=r(-D) \ .
\ee
The latter operator acts on functions of $z$  according to
the rule
\be
r(D)\cdot z^n=r(n)z^n \ .
\ee
 The exponents in (\ref{psixir}),(\ref{psi*xir}) are defined by
their Taylor series.

Using relations (\ref{psixir}),(\ref{psi*xir}) and the fact that
inside $res _z$ the operator $\frac 1z r(D)$ is the conjugate of
$\frac 1z r(-D)=\frac 1z r'(D)$, we get

\blemm \label{Lemma bilinidA} The fermionic operator $e^{A({\bf t^*})}$
satisfies the bilinear identity (\ref{bilinear}):
\be
\label{bilAA}
\left[ \textrm{res}_{z=0} \psi(z) \otimes \psi^*(z),e^{A({\bf t^*})}\otimes
e^{A({\bf t^*})}\right]=0 \ .
\ee
\elemm
This lemma follows from the general approach in \cite{DJKM}.

By (\ref{bilAA}), the expression
\be
\label{tauhyp1}
\tau_r(n,{\bf t},{\bf t^*}):=
\langle n|e^{H({\bf t})} e^{-A( {\bf t^* })} |n\rangle 
\ee
provides us, for each choice of $n$, $r$ and ${\bf t}^*$, with a KP
tau-function (\ref{taucorKP}).  For any given  choice of $r$, the function
$\tau_r(n,{\bf t},{\bf t^*})$, viewed as a function of all the variables, is a
$2D$ Toda lattice tau-function.
\bprop \label{Schurexpansion}
\be
\label{tauhyp}
\tau_r(n,{\bf t},{\bf t^*} )  =
\sum_{{\lambda}} r_{ \lambda}(n)s_{\lambda }({\bf t})s_{\lambda
}({\bf t^* })  \ ,
\ee
where the operator valued exponential $ e^{-A( {\bf t^* })}$ inside
the vacuum expectation value is defined by its Taylor series and
the r.h.s is viewed as a formal series. (The sum in (\ref{tauhyp}) should be
understood as including the zero partition.)
 \eprop

Note that the tau-function (\ref{tauhyp}) can be viewed as resulting from 
the action of additional symmetries \cite{D, OW} on the vacuum tau-function.
The variables ${\bf t}$ play the role of KP flow parameters, and ${\bf t^*}$ is
a set of group parameters defining the exponential of a subalgebra of additional
symmetries of KP (see \cite{OW}). Alternatively, (\ref{tauhyp})
may be viewed as a tau-function of the two-dimensional Toda
lattice \cite{T, UT} with two sets of continuous variables
${\bf t},{\bf t^*}$ and one discrete variable $n$. Formula
(\ref{tauhyp}) is obviously symmetric with
respect to ${\bf t} \leftrightarrow {\bf t^*}$. \\

\subsection{The generalized Cauchy-Littlewood identity and a proof of
Prop. \ref{Schurexpansion} }

What is meant here by the generalized Cauchy-Littlewood identity
is the following relation, which can be interpreted as a (double)
generating function for the Schur functions.
\be
\label{Schurdoublegener}
e^{\sum_{m=1}^\infty mt_mt_m^*}=\sum_\lambda s_\lambda({\bf
t})s_\lambda({\bf t}^*) \ .
\ee
When the parameters ${\bf t}=(t_1, t_2, \dots )$ and  ${\bf
t}^*=(t^*_1, t^*_2, \dots)$ are given by (\ref{finiteparam}) in
terms of two  finite sets of variables $(x_1, \dots x_N)$ and
$(y_1, \dots y_M)$, this reduces to the usual form (\ref{Cauchy})
of the Cauchy-Littlewood identity. This will be proved here using the fermionic
Fock space and scalar product formulae discussed above. (In the previous papers
\cite{Or, OS}  only sketches of these proofs were presented.)

For the proof, we  need a preliminary lemma \cite{DJKM}.

\blemm \label{orderlemma}
For $-j_1<\cdots <-j_k<0\le i_k<\cdots <i_1$ the following formula is
valid:
\bea
\label{glemma'}
\l 0|e^{H({\bf t})}\psi^*_{-j_1}\cdots
\psi^*_{-j_k}\psi_{i_k}\cdots\psi_{i_1}|0\r&=&\l n |e^{H({\bf
t})}\psi^*_{-j_1+n}\cdots \psi^*_{-j_k+n}\psi_{i_k+n}\cdots\psi_{i_1+n}|n\r
\cr
&=&  (-1)^{j_1+\cdots
+j_k}s_{{\lambda}}({\bf t}) ,
\eea
where the partition ${\lambda}=( \lambda_{1},\dots ,
\lambda_{j_1})$ is defined by
\begin{eqnarray}\label{lambdaFrob}
(\lambda_{1},\dots , \lambda_{j_1})=(i_{1},\dots , i_k|j_1-1,\dots
, j_k-1) .
\end{eqnarray}
 (Here $(\dots | \dots)$ is the Frobenius notation for a partition.) 
\elemm 
The proof of this lemma  follows from a direct calculation, with the help of the
Wick rule \cite{DJKM}.

We now introduce the vectors (cf. \cite{TI}) \be
\label{rlambdanfermion} |\lambda,n \r :=\psi^*_{-j_1+n}\cdots
\psi^*_{-j_k+n}\psi_{i_k+n}\cdots\psi_{i_1+n}|n\r \ee \be
\label{llambdanfermion} \l\lambda,n|  := \l
n|\psi^*_{i_1+n}\cdots\psi^*_{i_k+n}\psi_{-j_k+n}\cdots
\psi_{-j_1+n} \ . \ee
 It follows that
\be
\label{ortlambdan}
\l \lambda,n|\mu,m\r=\delta_{mn}\delta_{\lambda\mu}  \ ,
\ee so the set of vectors (\ref{rlambdanfermion}) and
(\ref{llambdanfermion}) form orthonormal bases for the Fock
space $F$ and its dual $F^*$, respectively.

For any partition $\lambda$ expressed in Frobenius form as (\ref{lambdaFrob}),
we define the integer
\begin{equation}\label{a(lambda)}
a(\lambda):=j_1+\cdots+j_k.
\end{equation}
Then Lemma  \ref{orderlemma} and the orthogonality
relations (\ref{ortlambdan}) imply the expansions
 \be \label{0} \l
n|e^{H({\bf t})}=\sum_\lambda (-1)^{a(\lambda)}s_\lambda({\bf
t})\l \lambda,n| 
\ee 
and
\be
\label{0*} e^{H^*({\bf t}^*)}|n\r=\sum_\lambda |\lambda,n\r
(-1)^{a(\lambda)}s_\lambda({\bf t}^*) \ . \ee

\goodbreak
\noindent
{\bf Proof of the generalized Cauchy-Littlewood identity.}

Consider the vacuum TL tau function
\be
\label{-2}
\l 0|e^{H({\bf t})}e^{H^*({\bf t}^*)}|0\r  \ .
\ee
By (\ref{0}),(\ref{0*}) this is equal to the r.h.s. of
(\ref{Schurdoublegener}):
\be
\label{-2'}
\l 0|e^{H({\bf t})}e^{H^*({\bf t}^*)}|0\r=\sum_\lambda
s_\lambda({\bf t})s_\lambda({\bf t}^*) \ ,
\ee
(where the sum again includes the zero partition).

It follows from the Heisenberg algebra commutation relation
(\ref{Heisenberg}) that conjugation of  $e^{H_{-m}}$  by $e^{H_m}$ just
translates the coefficient of the exponent by $m$. This, together
with  eqs. (\ref{Hvac}) implies that that the vacuum TL tau function (\ref{-2})
is equal to the l.h.s. of (\ref{Schurdoublegener}):
\be
\label{vacTLett}
\l 0|e^{H({\bf t})}e^{H^*({\bf t}^*)}|0\r=e^{\sum_{m=1}^\infty m t_mt^*_m} \ .
\ee

\noindent
{\bf Proof of Prop. \ref{Schurexpansion}}. Consider the Taylor series in all
time variables $t_1,t_2,\dots $ defined by
\be
\label{1}
e^{A({\bf t}^*)}|n\r=\sum_{n_1,n_2,\dots=0}^\infty
t_1^{n_1}t_2^{n_2}\cdots A_1^{n_1}A_2^{n_2}\cdots |n\r  \ .
\ee
Since the $A_k$'s all commute, the order of factors on the
 r.h.s. is irrelevant. Each $A_k$  has the structure
\be
\label{2} A_k=\sum_{m=-\infty}^\infty e_{m,m-k} \ ,
\ee where
\be
e_{m,m-i}:=\psi_m\psi_{m-i}^*r(m)r(m-1)\cdots r(m-i+1) \ .
\ee

Recall that
\be
\label{creat|n>}
\psi_m|n\r =0,\quad m<n,\qquad \psi_m^*|n\r=0,\quad m\ge n \ .
\ee

For a given $n$, it is convenient to decompose  ${\widehat
gl}(\infty)$ (as a linear space, not as a Lie algebra) into the
following direct sum 
\be
\label{split} a=a^{++}\oplus a^{--}\oplus
a^{+-}\oplus a^{-+},\quad a\in {\widehat gl}(\infty)  \ ,
\ee 
where
\begin{eqnarray}\label{splitsum} a^{++}=\sum_{i,k\ge
n}:\psi_i\psi_{k}^*:a_{ik},\quad
  a^{--}=\sum_{i,k< n}:\psi_i\psi_{k}^*:a_{ik},\\
  a^{+-}=\sum_{i\ge n,k< n}:\psi_i\psi_{k}^*:a_{ik},\quad
  a^{-+}=\sum_{i<n,k\ge n}:\psi_i\psi_{k}^*:a_{ik} \ .
\end{eqnarray}
It follows that
\be
\label{4} [e_{i,j}^{+-},e_{k,m}^{+-}]=0,\quad
[A_i^{+-},A_k^{+-}]=0 \ ,
\ee
while
\be
\label{5}
[e_{i,m}^{++},e_{m,k}^{+-}]=e_{i,k}^{+-},\quad
[e_{i,m}^{+-},e_{m,k}^{--}]=e_{i,k}^{+-}
\ee
Each term in the sum on the r.h.s. of (\ref{1}) is a linear
combination of monomials  of the form
\be
\label{6}
e_{i_1,k_1}\cdots e_{i_N,k_N}|n\r
\ee
for some $N$. The product contains terms
$e^{+-}_{i,k},e^{++}_{i,k},e^{--}_{i,k}$ (with various subindices
$i,k$) , where the last two annihilate the vacuum vector $|n\r$.
Using the commutation relations (\ref{5}) we put all elements
$e^{++}_{i,k},e^{--}_{i,k}$ to the right in (\ref{6}), then reduce
(\ref{6}) to  a product of only the $e^{+-}_{i,k}$ terms:
\be
\label{7}
e_{i_1,k_1}^{+-}\cdots e_{i_M,k_M}^{+-}|n\r, \quad M\le N  \ .
\ee

Now consider (for $i>0$)
\be
\label{10}
H_{-i}=\sum_{m=-\infty}^\infty E_{m,m-i} \ ,
\ee
where
\be
E_{m,m-i}:=\psi_m\psi_{m-i}^*  \ ,
\ee
in terms of which we may write
\be
\label{11}
e_{m,m-i}: =E_{m,m-i}r(m)r(m-1)\cdots r(m-i+1)
\ee
\be
\label{12}
e_{i_1,k_1}^{+-}\cdots e_{i_M,k_M}^{+-}|n\r=r_\lambda(n)
E_{i_1,k_1}^{+-}\cdots E_{i_M,k_M}^{+-}|n\r \ .
\ee
In the expression
\be
\label{6*}
E_{i_1,k_1}^{+-}\cdots E_{i_M,k_M}^{+-}|n\r  \ ,
\ee
note that $E_{i,k}^{+-}E_{j,m}^{+-}=E_{i,m}^{+-}E_{j,k}^{+-}$ and
therefore one can apply the action of the permutation group and
commutation relations (\ref{4}) to reorder the first indices in
decreasing order, i.e. $i_1\ge i_2 \ge \cdots $ and the second
indices such that $-k_1\ge -k_2 \ge \cdots $. Then \be\label{8}
e_{i_1,k_1}^{+-}\cdots e_{i_M,k_M}^{+-}|n\r=r_\lambda(n)
E_{i_1,k_1}^{+-}\cdots E_{i_M,k_M}^{+-}|n\r
=r_\lambda(n)|\lambda,n\r (-1)^{M}, \ee where the partition
$\lambda$ in Frobenius notation is \be\label{9} \lambda=(i_1,\dots
,i_M|-k_1-1,\dots,-k_M-1 )  \ . \ee Since from (\ref{0*}) we have
\be \label{13} e^{H^*({\bf t^*})}|n\r=\sum_\lambda |\lambda,n\r
(-1)^{a(\lambda)} s_\lambda({\bf t}^*) \ , \ee we finally get \be
\label{14} e^{A({\bf t^*})}|n\r=\sum_\lambda |\lambda,n\r
(-1)^{a(\lambda)}r_\lambda(n)s_\lambda({\bf t}^*)  \ , \ee which
proves (\ref{tauhyp}). From this we may also deduce the following
result. \bprop \cite{Or} \label{schurorthog} \be \label{tauscal}
<s_\lambda,s_\mu >_{r,n}=\l n |s_\lambda({\bf H})s_\mu({\bf -A})
|n \r \ , \ee where the variables $(\gamma_1,
\gamma_2 \dots)$ forming the arguments of the Schur functions (see
(\ref{Schurht})) are replaced by the operators \be \label{HA}
  {\bf H}:=(\frac{H_1}{1},\frac{H_2}{2},\frac{H_3}{3},\dots ),\quad {\rm and}
\quad
 {\bf -A}:=(-\frac{A_1}{1},-\frac{A_2}{2},-\frac{A_3}{3},\dots )  \ .
\ee
\eprop

\noindent
{\bf Proof}. From the generalized Cauchy-Littlewood
identity we have
\be
\label{0'}
\l n|e^{H({\bf t})}=\sum_\lambda s_\lambda({\bf t})\l n|
s_\lambda({\bf H}) ,\quad e^{A({\bf t^*})}|n\r=\sum_\mu s_\mu({\bf
-A})|n\r s_\mu({\bf t}^*)
\ee
It follows from  (\ref{14}),(\ref{0}) and (\ref{ortlambdan}) that formula
 (\ref{0'}) is equivalent to (\ref{tauscall}), given in the following
proposition, which allows us to express $\tau_r(n,{\bf t},{\bf t^*})$ in terms
of the scalar product $\l  \ , \ \r  _{r,n}$.

\bprop \cite{Or}  \label{tauscalarprod}
\be
\label{tauscall}
\tau_r(n,{\bf t},{\bf t^*})=\l e^{\sum_{m=1}^\infty
mt_m\gamma_m},e^{\sum_{m=1}^\infty mt_m^*\gamma_m}\r  _{r,n} \ ,
\ee
 where we consider the scalar product on the space of  functions of
variables $\gamma$, while ${\bf t},{\bf t}^*$ are viewed as
parameters.
\eprop

\section{Matrix models}

\subsection{Normal matrix model.}

 We  now consider an ensemble of $n \times n$  normal matrices $M$ (i.e.,
matrices which commute with their Hermitian conjugates). The following
integral defines the partition function for the  normal matrix model.
\bea
I^{NM}(n,{\bf t},{\bf t^*};{\bf u})&=&\int dMdM^+
e^{\tr(V_1( M)+V_2( M^+)-MM^+)} \label{normint} \cr
&=&
C_n\int_{\texttt c} \cdots \int_{\texttt c} |\Delta({\bf z})|^2
e^{\sum_{i=1}^n(V_1(z_i)+V_2({\bar z}_i)-z_i{\bar z}_i)}\prod_{i=1}^n dz_id{\bar
z}_i \ ,  \label{NMPF}
\eea
where
\be
dM=\prod_{i<k} d\Re M_{ik} d\Im M_{ik}\prod_{i=1}^n dM_{ii} \ .
\ee
(This model has applications, e.g., to the Laplacian growth problem
\cite{MWZ}.) The last integral in (\ref{NMPF}) is taken over $n$ copies of the 
complex plane,  $(z_1, \dots z_n)$ are the eigenvalues of the matrix $M$ and
$C_n$ is a normalization factor coming from  the matrix angular integration.  
Here $V_1$ and $V_2$ are the power series \be V_1(z):=\sum_{m=1}^\infty t_m z^m,
\qquad V_2(z):=\sum_{m=1}^\infty t_m^* z^m \ . \ee

 Formula (\ref{realizA}) then yields a series expansion  of
the integral (\ref{normint}) as:
\be
\label{INMpert}
I^{NM}(n,{\bf t},{\bf t^*};{\bf
u})=<e^{\sum_{i=1}^n\sum_{m=1}^\infty t_mz_i^m}
,e^{\sum_{i=1}^n\sum_{m=1}^\infty t_m^*{ z}_i^m}
>_{n}'=\sum_{\lambda}(n)_\lambda s_\lambda({\bf
t})s_\lambda({\bf t^*}) \ .
\ee
This formula may be interpreted \cite{HO1, HO2} as a multi-dimensional
analog of the Borel summation of the series on the RHS which, generally,
is divergent.

\subsection{Two-matrix model.}
Let us evaluate the following partition function defined on an ensemble
consisting of pairs $(M_1, M_2)$ of $n \times n$ Hermitian matrices $M_1$ and 
$n \times n$ anti-Hermitian matrices $M_2$. 
\be 
\label{HTMMmeasure} I^{2MM}(n,{\bf t},{\bf t^*})=\int
e^{\tr(V_1(M_1)+V_2(M_2)-  M_1M_2)}dM_1dM_2 \ . 
\ee 
It is well-known \cite{Mehta, HCIZ} that this integral reduces to
the following one over the eigenvalues $x_i$ and $y_i$  of the
matrices $M_1$ and $M_2$ respectively: 
\be
 {\tilde C}_n\int_\Re\int_\Im
\cdots \int_\Re \int_\Im e^{\sum_{j=1}^n(\sum_{m=1}^\infty (t_m
x_j^m + t^*_m y_j^m)  - x_jy_j)}
 {\Delta({\bf x})\Delta({\bf y})}\prod_{j=1}^n
dx_jdy_j  \ .
\ee
with the normalization factor ${\tilde C}_n$ proportional to the unitary
group volume. It was shown in \cite{GMMMO} that this integral is a
two dimensional TL tau-function.

From (\ref{realizB}),  we obtain that the partition function for
these random matrices is a series of hypergeometric type
\be
\label{2MM=scalprod}
I^{2MM}(n,{\bf t},{\bf t^*})=\frac{{\tilde C}_n}{(2\pi i)^n}
<e^{\sum_{i=1}^n\sum_{m=1}^\infty t_mz_i^m} ,e^{\sum_{i=1}^n\sum_{m=1}^\infty
t_m^*{ z}_i^m} >_{n}'=\frac{{\tilde C}_n}{(2\pi i)^n}\sum_{\lambda}(n)_\lambda
s_\lambda({\bf t})s_\lambda({\bf t^*}) \ ,
\ee
where we use the fact that, according to (\ref{rlambda})
\be
\label{(n)lambda}
(n)_\lambda=\prod_{i,j\in
\lambda}(n+j-i)=\frac{\Gamma(n+1+\lambda_1)\Gamma(n+\lambda_2)\cdots
\Gamma(\lambda_n)}{\Gamma(n+1)\Gamma(n)\cdots \Gamma(1)} \ .
\ee

We thus have the following perturbation series
\be
\label{2MMss}
\frac{I^{2MM}(n,{\bf t},{\bf t^*})}{I^{2MM}(n,0,0)}
=\sum_{\lambda} (n)_{\lambda}s_{\lambda}({\bf t})s_{\lambda}({\bf
t}^*) \ .
\ee

When all higher parameters $t_j, t^*_j$ with $ j >2$  vanish (i.e,
the case of a Gaussian matrix integral) and $n=1$, this series can
easily be evaluated as (cf. \cite{HO2})
\be
\label{gauss2MM}
I^{2MM}(1,t_1,t_2,0,0,\dots;t_1^*,t_2^*,0,0,\dots)=\sum_{m=0}^\infty
m!h_m({\bf t})h_m({\bf t^*})= \frac {\exp
\frac{t_1t_1^*+t_2\left(t_1^*\right)^2+t_2^*(t_1)^2}
{1-4t_2t_2^*}}{\sqrt{1-4t_2t_2^*}} \ .
\ee

\subsection{Hermitian one-matrix model.}
It was shown in \cite{GMMMO} that the partition function for the Hermitian
one-matrix model model is a one dimensional Toda tau-function that satisfies
the so-called Virasoro constraints. Here we consider the perturbation series
for this model expressed in terms of Schur functions. Let
 $M$ be a Hermitian $N \times N$ matrix.

The partition function for the Hermitian one-matrix model with an even quartic
potential is
\be
\label{1HOMM}
Z(N,g,g_4)=\int dM e^{-N\tr(\frac g2 M^2+g_4 M^4)} \ .
\label{1MPF}
\ee
We choose the normalization of the matrix integral in such a way
that $Z(N,g,g_4)$ is equal to $1$ when $g_4=0$. Let us evaluate the simplest
perturbation terms for the one matrix model using the series
(\ref{gauss2MM}) \cite{HO2}. First,  consider the partition function
(\ref{HTMMmeasure}) for the two-matrix model and set all $t_k=0$
except $t_4$, and all $t_k^*=0$ except $t_2^*$. Then (cf.
\cite{BEH, Eyn}) it is known that if we put
\be
\label{tg}
g_4=-4N^{-1}t_4,\quad g=-(2Nt_2^*)^{-1} \ ,
\ee
we obtain
\be
\label{2MM=1MM}
I^{2MM}(0,0,0,t_4,0,\dots;0,t_2^*,0,0,\dots)=Z(N,g,g_4)  \ ,
\ee
 and therefore
\be
\label{1MM}
\sum_{\lambda} (N)_{\lambda}s_{\lambda}({\bf t})s_{\lambda}({\bf
t}^*)=\int dM e^{-N\tr(\frac g2 M^2+g_4 M^4)}=Z(N,g,g_4) \ .
\ee

\noindent
{\bf Remark}. It is well-known (see for instance
\cite{Eyn}) that, according to the Feynman rules for the one matrix
model: (a) to each propagator (double line) is associated a
factor $1/(Ng)$ (which is $-2t_2^*$ in our notations) (b) to each
four legged vertex is associated a factor $(-Ng_4)$ (which is $4t_4$
in our notations) (c) to each closed single line is associated a
factor $N$. Therefore we may say that the factor $(N)_{\lambda}$
in (\ref{1MM}) is responsible for closed lines, the factors
$s_{\lambda}({\bf t}=0,t_2,0,\dots)$ are responsible for
propagators and the factors $s_{\lambda}({\bf t}^*)$ are
responsible for vertices. It follows that Feynman diagrams containing
$k=|{\lambda}|/4$ vertices and $2k=|{\lambda}|/2$ propagators give the terms:
\be
\label{numberN}
(Ng_4)^{|{\lambda}|/4}(Ng)^{-|{\lambda}|/2}\sum_{|{\lambda}|=4k}
(N)_{\lambda}s_{\lambda}(0,0,0,1,0,\dots)s_{\lambda}(0,1,0,\dots)
\ ,
\ee
where the number $|{\lambda}|$ is the weight of partition
${\lambda}$.

An  explicit calculation of the first three nonvanishing terms 
(corresponding  to partitions of weight $0$, $4$ and $8$) yields the
same result as the Feynman graph calculation (cf. \cite{Eyn})
\be
\label{pertser1M}
Z(N,g,g_4)=1-\frac{g_4}{g^2}\left(\frac{N^2}{2}+\frac 14
\right)+\frac{g_4^2}{g^4}(32N^4+320N^2+488)+
O\left(\frac{g_4^3}{g^6} \right)  \ .
\ee

\section{Expression for $\log \tau_r(n,{\bf t},{\bf t}^*)$}

Recall that a product of two Schur functions can be expanded as a
sum \cite{Mac}
\be
\label{RL}
s_\lambda({\bf t})s_\mu({\bf
t})=\sum_{\nu}C^\nu_{\lambda\mu}s_\nu({\bf t})  \ ,
\ee
where the coefficients $C^\nu_{\lambda\mu}$ may be calculated via the
Littlewood-Richardson combinatorial rule \cite{Mac}, and the weight $|\nu|$ of
the partitions in the sum is equal to the sum of the weight:
\be
\label{weights}
|\nu|=|\lambda|+|\mu|  \ .
\ee
Writing (\ref{tauhyp}) as $1+\sum_{\lambda \neq
 0}r_\lambda(n)
 s_\lambda({\bf t}) s_\lambda({\bf t}^*) $
 and taking the log, we obtain
\be
\label{logtau}
\log \tau_r(n,{\bf t},{\bf t}^*)=\sum_{\lambda,\lambda'\neq 0}
s_\lambda({\bf t})k_{\lambda\lambda'}(n)s_{\lambda'}({\bf t}^*)  \ ,
\ee
where $k_{\lambda\lambda'}$ is given by the following series
\be
\label{klambdalambda}
k_{\lambda\lambda'}(n)=\frac{1}{1}r_\lambda(n)\delta_{\lambda\lambda'}+
\sum_{k>1}\frac{1}{k}\sum_{\lambda_1,\dots,\lambda_k \neq 0}
r_{\lambda_1}(n)\cdots r_{\lambda_k}(n)
\sum_{\nu_1,\dots,\nu_{k-2}}
C^\lambda_{\nu_1,\dots,\nu_{k-2}}\sum_{\nu_1',\dots,\nu_{k-2}'}
C^{\lambda'}_{\nu_1',\dots,\nu_{k-2}'}  \ ,
\ee
with
\bea\label{C}
C^\lambda_{\nu_1,\dots,\nu_{k-2}}&:=&C_{\lambda_1\lambda_2}^{\nu_1}
C_{\nu_1\lambda_3}^{\nu_2}\cdots
C_{\nu_{k-3}\lambda_{k-1}}^{\nu_{k-2}}C_{\nu_{k-2}\lambda_k}^{\lambda}  \ ,
\cr
C^{\lambda'}_{\nu_1',\dots,\nu_{k-2}'}&:=&
C_{\lambda_1\lambda_2}^{\nu_1'} C_{\nu_1'\lambda_3}^{\nu_2'}\cdots
C_{\nu_{k-3}'\lambda_{k-1}}^{\nu_{k-2}'}C_{\nu_{k-2}'\lambda_k}^{\lambda'} \ .
\eea
The first few terms are
\bea
k_{\lambda\lambda'}(n)=&&\frac{1}{1}r_\lambda(n)\delta_{\lambda\lambda'}+
\frac{1}{2}\sum_{\lambda_1,\lambda_2\neq
0}r_{\lambda_1}(n)r_{\lambda_2}(n)
 C_{\lambda_1\lambda_2}^\lambda
C_{\lambda_1\lambda_2}^{\lambda'}
\cr
&&+\frac{1}{3}\sum_{\lambda_1,\lambda_2,\lambda_3\neq
0}r_{\lambda_1}(n)r_{\lambda_2}(n) r_{\lambda_3}(n)\sum_{\nu}
C_{\lambda_1\lambda_2}^\nu C_{\nu\lambda_3}^{\lambda}\sum_{\nu'}
C_{\lambda_1\lambda_2}^{\nu'} C_{\nu'\lambda_3}^{\lambda'}+\cdots
\eea
Note that, due to (\ref{weights}), each
$k_{\lambda\lambda'}(n)$ consists of a sum over just a finite number of terms.


\begin{thebibliography}{99}


\bibitem{BEH}  M. Bertola, B. Eynard and J. Harnad, ``Partition functions for
Matrix Models and Isomonodromic Tau functions'', {\it J. Phys. A. Math, Gen.}
{\bf 6} (2003, in press), nlin.SI/0204054.

\bibitem{DJKM} E. Date, M. Jimbo, M. Kashiwara and T. Miwa,
``Transformation groups for soliton equations'', in: {\it  Nonlinear
integrable systems -- classical theory and quantum theory}, (eds. M.
Jimbo, and T. Miwa, World Scientific, 39--120, 1983.

\bibitem{D} L. A. Dickey,  {\em Soliton Equations and Hamiltonian Systems},
World Scientific, Advanced Series in Mathematical Physics Vol. {\bf 12},
Singapore, New Jersey, London (1993).

\bibitem{Eyn} B. Eynard,  ``Random Matrices'', Cours de Physique Theorique de
Saclay, preprint Saclay-T01/014, CRM-2708 (2001).

\bibitem{VF} M. Feigin and A. P. Veselov, ``Quasiinvariants of
Coxeter groups and $m$-harmonic polynomials'', {\it Intern. Math. Res. Notices}
{\bf 10}, 521-545 (2002).

\bibitem{GMMMO}  A. Gerasimov, A. Marshakov, A. Mironov, A. Morozov
and A. Orlov, ``Matrix Models of 2D Gravity and Toda Theory'',
Nuclear Physics {\bf B357}  565-618 (1991).

\bibitem{HO1}
J. Harnad and A. Yu.  Orlov, ``Matrix integrals as Borel sums of Schur
function expansions'',  preprint CRM-2865 (2002), nlin.SI/0209035.
To appear in: Symmetries and Perturbation theory SPT2002, eds. S. Abenda and
G. Gaeta, (World Scientific, Singapore), 2002/3 (in press)).

\bibitem{HO2} J. Harnad and A. Yu. Orlov, ``Schur Function Expansions of Matrix
Integrals'', preprint CRM December 2002.

\bibitem{HCIZ} C. Itzykson and J.B. Zuber, ``The planar approximation II'',
 {\it J. Math. Phys.} {\bf 21},  411-421  (1980).

\bibitem{Mac} I. G. Macdonald, Symmetric Functions and Hall Polynomials,
Clarendon Press, Oxford (1995).

\bibitem{Mehta} M. L. Mehta, {\em Random Matrices} 2nd ed. , Academic Press,
 San Diego (1991).

\bibitem{MWZ} M. Mineev-Weinstein, P. Wiegmann and A. Zabrodin,
``Integrable structure of interface dynamics'', {\it Phys. Rev. Lett.}
{\bf 84}, 5106-5109 (2000).

\bibitem{Or} A. Yu. Orlov, ``Soliton theory, Symmetric Functions
and Matrix Integrals'',  preprint (2002), SI/0207030.

\bibitem{OS} A. Yu. Orlov and D. M. Scherbin ``Hypergeometric solutions
of soliton equations'', {\it Theor. and Math. Phys.} {\bf 128}, 84-108 (2001).

\bibitem{OW} A. Yu. Orlov and P. Winternitz,
``$P_\infty$ symmetries of integrable systems'', {\it Theor. and Math.
Phys.}, {\bf 113}, 1393-1417 (1997).

\bibitem{T} K. Takasaki, ``The Toda Lattice Hierarchy and
Generalized String Equation'', {\it Commun. Math. Phys.} {\bf  181}  131
(1996).

\bibitem{TI} T. Takebe, ``Representation Theoretical Meaning
of Initial Value Problem for the Toda Lattice Hierarchy I'',
{\it Lett. Math. Phys.} {\bf 21}: 77--84 (1991).

\bibitem{UT} K. Ueno and K. Takasaki,
``Toda Lattice Hierarchy'', {\it Adv. Studies Pure Math.} {\bf 4}, 1--95
(1984).

\bibitem{ZM} V.E. Zakharov,  S.V. Manakov,   S.P. Novikov, and  L.P. Pitaevsky,
{\em Theory of Solitons. The Inverse Scattering Method}, Nauka, Moscow (1980);
English transl.: Contemporary Sov. Math., Consultants Bureau (Plenum), New
York, London (1984).

\end{thebibliography}
\end{document}